\documentclass[11pt,onecolumn,amssymb,nofootinbib]{revtex4}
\usepackage{amsmath, amsthm, amscd, amssymb}
\usepackage{bm}
\usepackage{bbm}

\begin{document}

\title{\bf Implications of a frame dependent gravitational effective action for perturbations on the
Robertson-Walker Metric}

\author{Stephen L. Adler}
\email{adler@ias.edu} \affiliation{Institute for Advanced Study,
Einstein Drive, Princeton, NJ 08540, USA.}

\begin{abstract}
In earlier work we showed that a frame dependent effective action motivated by the postulates of three-space
general coordinate invariance and Weyl scaling invariance exactly mimics a cosmological constant in Robertson-Walker (RW)
spacetimes.  Here we study the implications of this effective action for small fluctuations around a spatially flat RW background
geometry. The equations for the conserving extension of the modified stress-energy tensor can be integrated in closed
form, and involve only the metric perturbation $h_{00}$.  Hence the equations for tensor and vector perturbations are
unmodified, but there are Hubble scale additions to the scalar perturbation equations, which nonetheless admit no
propagating wave solutions.  Consequently, there are
no modifications to standard gravitational wave propagation theory, but there may be observable implications for cosmology.  We give a self-contained discussion,
including an analysis of the restricted class of gauge transformations that act when a frame dependent effective action is present.

\end{abstract}

\maketitle

\section{Introduction and summary}

The experimental observation of an accelerated expansion of the universe has been interpreted as evidence for a cosmological
term in the gravitational action of the usual form
\begin{equation}\label{usual}
 S_{\rm cosm}=-\frac{\Lambda}{8 \pi G} \int d^4x (^{(4)}g)^{1/2}~~~,
\end{equation}
with $\Lambda=3H_0^2\Omega_{\Lambda}$  in terms of the Hubble constant $H_0$ and the cosmological fraction $\Omega_{\Lambda}\simeq 0.72$.
This functional form incorporates the usual assumption that gravitational physics is four-space general coordinate invariant,
with no frame dependence in the fundamental action.

In a series of papers \cite{adler1}-\cite{adler3}, motivated by the frame dependence of the cosmological background radiation,  we have studied the implications of the assumption that there is an induced
gravitation effective action that is three-space general coordinate and Weyl scaling invariant, but is not four-space general coordinate invariant.   For the special class of diagonal metrics for which $g_{0i}=0$, these assumptions imply that the term in the induced effective action with no
 metric derivatives has the  form
\begin{equation}\label{effact2}
 S_{\rm eff}=A_0 \int d^4x (^{(4)}g)^{1/2} (g_{00})^{-2}~~~,
\end{equation}
with $A_0$ a constant.
When $A_0$ is given the value
\begin{equation}\label{a0}
A_0=-\frac{\Lambda}{8\pi G}~~~,
\end{equation}
this effective action becomes
\begin{equation}\label{effact3}
 S_{\rm eff}=-\frac{\Lambda}{8\pi G} \int d^4x (^{(4)}g)^{1/2} (g_{00})^{-2}~~~,
\end{equation}
and in Robertson-Walker (RW) spacetimes where $g_{00}=1$ exactly mimics the cosmological constant
effective action of Eq. \eqref{usual}.

The  paper \cite{adler2} studied the implications of the effective action of Eq. \eqref{effact3} for spherically symmetric solutions of
the Einstein equations, and showed that in a static, spherically symmetric Schwarzschild-like geometry it modifies the black hole horizon structure within microscopic distances of the nominal horizon, in such a way that $g_{00}$ never vanishes.  This could have important implications,
yet to be investigated, for the black hole ``information paradox''.  In the present  paper we turn to studying the implications of the effective action
of Eq. \eqref{effact3} for the equations governing small perturbations around a spatially flat RW geometry.  We find that the equations for tensor perturbations governing gravitational waves are unchanged, as are the equations for vector perturbations.  However, the equations governing scalar perturbations receive Hubble scale
corrections, which could have implications, again yet to be investigated, for structure formation in the early universe.

To set up a phenomenology for testing for the difference between the cosmological actions of Eq. \eqref{usual} and Eq. \eqref{effact3}, we make
the Ansatz that the observed cosmological constant arises from a linear combination of these two actions of the form
\begin{equation} \label{ansatz}
S_{\Lambda} = (1-f)  S_{\rm cosm}+ f  S_{\rm eff}
= -\frac{\Lambda}{8\pi G} \int d^4x (^{(4)}g)^{1/2} [1-f + f(g_{00})^{-2}]~~~~,
\end{equation}
so that $f=0$ corresponds to only a standard cosmological constant, and $f=1$ corresponds to only an apparent cosmological constant
arising from a frame dependent effective action.  Our results for the modifications to the equations governing scalar perturbations will
thus contain the parameter $f$.

As noted in \cite{adler1} and again in \cite{adler3},  while the Einstein-Hilbert action and
the particulate matter action  are four-space general coordinate invariant, the frame-dependent effective action $ S_{\rm eff}$ is
invariant only under the subset of general coordinate transformations that act on the spatial coordinates $\vec x$, while leaving
the time coordinate $t$ invariant.  Consequently, the stress-energy tensor obtained by varying $ S_{\Lambda}$ with respect
to the full metric $g_{\mu\nu}$  will not satisfy the covariant conservation condition, and thus cannot be used as a source
for the full spacetime Einstein equations.  However, it is consistent to include $ S_{\Lambda}$ in the source for the spatial
components of the Einstein equations in the preferred rest frame of the action of Eq. \eqref{effact3}, which we identify
as the rest frame of the cosmological background radiation, giving the following rules:

\begin{enumerate}
\item  The spatial components of the Einstein equations are obtained by varying the full action with respect to $g_{ij}$, giving
\begin{equation}\label{einstein}
G^{ij}+8\pi G ( T_{\Lambda}^{ij} + T^{ij}_{\rm pm})=0  ~~~,
\end{equation}
with $T^{ij}_{\rm pm}$ the spatial components of the usual particulate matter stress-energy tensor, and with
$ T_{\Lambda}^{ij}$ given by
\begin{equation}\label{deltat}
\delta S_{\Lambda }=-\frac{1}{2} \int d^4x (^{(4)}g)^{1/2}  T_{\Lambda}^{ij} \delta g_{ij}~~~.
\end{equation}
\item  The components of the Einstein tensor $G^{0i}=G^{i0}$ and $G^{00}$ are obtained from the
Bianchi identities with $G^{ij}$ as input, and from them we can infer the conserving extensions $ T_{\Lambda}^{i0}$ and
$ T_{\Lambda}^{00}$ of the spatial stress-energy tensor components  $ T_{\Lambda}^{ij}$.
\end{enumerate}

Equivalently, we can infer these conserving extensions by imposing
the covariant conservation condition on the tensor $ T_{\Lambda}^{\mu \nu}$,  with $ T_{\Lambda}^{ij}$ as input, and this is
how we proceed in Section 2.  In Sec. 3 we analyze the residual gauge invariance when there is a frame dependent effective action.  In Sec. 4
we give the modified equations for the scalar perturbations, and in Sec. 5 prove that they do not lead to scalar wave propagation. Sec. 6
gives a brief conclusion.   In Appendix
A we summarize our notational conventions, and give their relation to those of the text of Weinberg \cite{wein}.  In Appendix B we give the
formulas for the zeroth and first order inverse metric, affine connections, and Ricci tensor components. In  Appendix C we show the equivalence between
two different forms of the gauge invariance constraint on the matter perturbations, and derive formulas used in setting up the scalar perturbation equations.  In Appendix D we show that our dark energy model cannot be tested  using  the effective field theory framework of Gubitosi et al. \cite{gub}.
In Appendix E, we show that using the more general effective action that applies when the metric is not diagonal (as is the case for RW perturbations) does not alter the first order perturbation equations.

\section{Conserving extension $ T_{\Lambda}^{\mu \nu}$ of $ T_{\Lambda}^{ij}$}

Adding a small perturbation $h_{\mu\nu}$ to the zeroth order spatially flat RW metric, we have for the total metric $g_{\mu\nu}$
\begin{align}\label{totalmetric}
g_{00}=&1+h_{00}~~~,\cr
g_{i0}=&g_{0i}=h_{i0}~~~,\cr
g_{ij}=&-a^2(t)\delta_{ij}+h_{ij}~~~.\cr
\end{align}
The inverse metric, affine connection, and Ricci tensor corresponding to Eq. \eqref{totalmetric}  are given through  first order terms
in $h_{\mu\nu}$ in Appendix B.

Varying the spatial metric components $g_{ij}$, we find from Eqs. \eqref{deltat} and \eqref{ansatz} that
\begin{align}\label{vary1}
T_{\Lambda}^{ij}= &\frac{\Lambda}{8\pi G}[(1-f) g^{ij}+ f \frac{g^{ij}}{g_{00}^2}] \cr
=&\frac{\Lambda}{8\pi G}[(1-f) g^{ij}+ f \frac{g^{ij}}{(1+h_{00})^2}]~~~.\cr
\end{align}
Expanding through first order in $h_{00}$ and using $g^{ij}=-\delta_{ij}/a^2(t)$ in the term proportional to $h_{00}$, this becomes
\begin{align}\label{vary2}
T_{\Lambda}^{ij}=& \frac{\Lambda}{8\pi G}[ g^{ij}+f t^{ij}]~~~,\cr
t^{ij}=&\frac{2\delta_{ij} h_{00}}{a^2(t)}~~~.
\end{align}
Since the metric is covariantly conserved, $D_{\nu}g^{\mu\nu}=0$, with both $D_{\mu}$ and
$g^{\mu\nu}$ accurate to first order in the perturbation $h_{\mu\nu}$,  the conserving extension of $g^{ij}$ is  $g^{\mu\nu}$, including both zeroth and
first order terms.  Thus our task is to find a conserving extension $t^{\mu\nu}$ of $t^{ij}$ that obeys the covariant conservation condition
$D_{\nu}t^{\mu\nu}=0$. Since $t^{ij}$ is already first order in the perturbation $h_{00}$, it suffices to solve this equation using the zeroth
order covariant derivative constructed by using the affine connections of Eq. \eqref{affineconn} with the first order perturbation terms
omitted.

Expanding $D_{\nu}t^{\mu\nu}=\partial_{\nu}t^{\mu\nu}+\Gamma^{\mu}_{\nu\alpha}t^{\alpha\nu}+\Gamma^{\nu}_{\nu\alpha}t^{\mu\alpha}$ in terms of
temporal and spatial index values, we get
\begin{align} \label{covcons}
0=&D_{\nu}t^{\ell\nu}=\partial_0 t^{\ell0}+\partial_jt^{\ell j}+ \Gamma^{\ell}_{00}t^{00}+2\Gamma^{\ell}_{0j}T^{0j}+\Gamma^{\ell}_{mn}t^{mn}
+\Gamma^{\nu}_{\nu 0}t^{\ell 0}+\Gamma^{\nu}_{\nu j}t^{\ell j}~~~,\cr
0=&D_{\nu}t^{0\nu}=\partial_0 t^{00}+\partial_jt^{0 j}+ \Gamma^{0}_{00}t^{00}+2\Gamma^0_{0j}T^{0j}+\Gamma^{0}_{mn}t^{mn}
+\Gamma^{\nu}_{\nu 0}t^{0 0}+\Gamma^{\nu}_{\nu j}t^{0 j}~~~,\cr
\end{align}
which on substituting the zeroth order affine connections from  Eq. \eqref{affineconn} becomes
\begin{align} \label{covcons1}
0=&D_{\nu}t^{\ell\nu}=\partial_0 t^{\ell 0}+\partial_jt^{\ell j}+2\frac{\dot{a}}{a}t^{0\ell}+3\frac{\dot{a}}{a}t^{\ell 0}\cr
=&\partial_0 t^{\ell 0}+\partial_jt^{\ell j}+5\frac{\dot{a}}{a}t^{0\ell}~~~,\cr
0=&D_{\nu}t^{0\nu}=\partial_0 t^{00}+\partial_jt^{0 j}+ a\dot{a}t^{mm}+3\frac{\dot{a}}{a}t^{00} ~~~.\cr
\end{align}

These differential equations are readily integrated, to give
\begin{align}\label{integrate}
t^{ij}(\vec x,t)=&2\delta_{ij}a^{-2}(t)h_{00}(\vec x,t)~~~,\cr
t^{\ell 0}(\vec x, t)=&-2a^{-5}(t)\int_{L_1}^t du a^3(u)  \partial_{\ell}h_{00}(\vec x,u) ~~~,\cr
t^{00}(\vec x,t)=&-a^{-3}(t)\int_{L_2}^t du a^3(u) \left[\partial_{\ell}t^{0\ell}(\vec x, u)+6 \frac{\dot{a}(u)}{a(u)}h_{00}(\vec x, u)\right]~~~.
\end{align}
The lower limits $L_1$ and $L_2$ are arbitrary constants of integration; if we add an initial condition that all perturbations
should be bounded at the initial time $t_{\rm init}$ where $a(t_{\rm init})=0$,  then we should take $L_1=L_2= t_{\rm init}$.

To write the Einstein equations in terms of the Ricci tensor, as in Eq. \eqref{ricciform}, we need $t_{ij}$, $t_{i0}$, $t_{00}$, and
the trace $t_\alpha^\alpha$.  These can be written as (taking the integration limits now as $t_{\rm init})$
\begin{align}\label{integrate1}
t_{ij}(\vec x,t)=&2\delta_{ij}a^{2}(t)h_{00}(\vec x,t)~~~,\cr
t_{\ell 0}(\vec x, t)=&t_{0 \ell}(\vec x, t)=2a^{-3}(t)\int_{t_{\rm init}}^t du a^3(u)  \partial_{\ell}h_{00}(\vec x,u) ~~~,\cr
t_{00}(\vec x,t)=&a^{-3}(t)\int_{t_{\rm init}}^t du  [a(u)\partial_{\ell}t_{0\ell}(\vec x, u)-6 a^2(u)\dot{a}(u)h_{00}(\vec x, u)]~~~,\cr
\partial_{\ell}t_{\ell 0}(\vec x,t)=&2a^{-3}(t)\int_{t_{\rm init}}^t du a^3(u) \nabla^2 h_{00}(\vec x, u)~~~,\cr
t_{\alpha}^{\alpha}(\vec x, t)=& t_{00}(\vec x, t)-6h_{00}(\vec x,t)~~~.\cr
\end{align}
Substituting these equations into Eq. \eqref{ricciform} gives our final result for the modified Einstein equations, in Ricci tensor form.

Let us now separate all terms of Eq. \eqref{ricciform} into zeroth and first order parts.  Writing
\begin{equation}\label{riccisep}
R_{\mu\nu}=R_{\mu\nu}^{(0)}+R_{\mu\nu}^{(1)}~~~,
\end{equation}
we read off from Eq. \eqref{riccitensor} that
\begin{align}\label{riccisep1}
R_{00}^{(0)}=&3 \frac{\ddot{a}}{a}~~~,\cr
R_{0i}^{(0)}=&0~~~,\cr
R_{ij}^{(0)}=&-\delta_{ij}[a\ddot{a}+ 2(\dot{a})^2]~~~.\cr
\end{align}
We make a similar splitting for $T_{\mu \nu{\rm pm}}$,  taking the zeroth order stress energy
tensor to be
\begin{align}\label{zerostress}
T^{(0)}_{\mu \nu{\rm pm}}=&(p+\rho)u_{\mu}u_{\nu}-pg_{\mu \nu}~~~,\cr
T^{\alpha(0)}_{\alpha{\rm pm}}=&\rho-3p~~~,\cr
\end{align}
with $\rho$ and $p$ the particulate matter
density and pressure, and $u_{\mu}$ the four velocity with $u_{0}=1,\, u_{i}=0$. The zeroth order part of Eq. \eqref{ricciform} gives the standard
equations governing RW cosmology,
\begin{align}\label{zerothordercosm}
\frac{\ddot{a}}{a}+ 2(\frac{\dot{a}}{a})^2=&\Lambda +8\pi G\frac{1}{2}(\rho-p)~~~,\cr
3 \frac{\ddot{a}}{a}=&\Lambda-8 \pi G \frac{1}{2}(\rho+3p)~~~.\cr
\end{align}
The first order part of Eq. \eqref{ricciform} is
\begin{equation}\label{firstorder}
R^{(1)}_{\mu \nu}-\Lambda h_{\mu\nu}=-8 \pi G [T_{\mu \nu{\rm pm}}-\frac{1}{2} g_{\mu\nu} T^\alpha_{\alpha {\rm pm}}]^{(1)}
-\Lambda f[ t_{\mu \nu}-\frac{1}{2} g^{(0)}_{\mu\nu}  t^\alpha_{\alpha }]~~~,
\end{equation}
with
\begin{equation}\label{combo}
[T_{\mu \nu{\rm pm}}-\frac{1}{2} g_{\mu\nu} T^\alpha_{\alpha {\rm pm}}]^{(1)}=
T^{(1)}_{\mu \nu{\rm pm}}-\frac{1}{2}[ g^{(0)}_{\mu\nu} T^{\alpha(1)}_{\alpha {\rm pm}}+h_{\mu\nu}  T^{\alpha(0)}_{\alpha {\rm pm}}]~~~.
\end{equation}
Rewriting Eq. \eqref{firstorder} with all terms on the same side of the equation,
it is
\begin{equation}\label{firstorder1}
0= R^{(1)}_{\mu \nu}-\Lambda h_{\mu\nu}+8 \pi G [T_{\mu \nu{\rm pm}}-\frac{1}{2} g_{\mu\nu} T^\alpha_{\alpha {\rm pm}}]^{(1)}
+\Lambda f[ t_{\mu \nu}-\frac{1}{2} g^{(0)}_{\mu\nu}  t^\alpha_{\alpha }]~~~.
\end{equation}

\section{Residual gauge invariance with a frame dependent effective action}

Since the effective action of Eq. \eqref{effact3} is not four-space general coordinate invariant, but only three-space invariant,
the gauge invariance group of the first order perturbation equations will be reduced.  Let us consider the infinitesimal transformation
\begin{equation}\label{gauge1}
x^{\alpha}=x^{\prime\alpha}-\epsilon^{\alpha}(x^{\prime})~~~,
\end{equation}
with $\epsilon^0=0$, so that $t=t^{\prime}$.\footnote{With a nonzero $\epsilon^0(\vec x)$ that is independent of $t$, $\delta_gh_{00}$ remains zero \big(see \cite{wein}, Eq. (5.3.7)\big).   This additional gauge invariance can be used to impose a condition at only one time, and we do not use
  it in what follows.}  The metric tensor tensor transforms according to
\begin{align}\label{gauge2}
g^{\prime}_{\mu^{\prime}\nu^{\prime}}(x^{\prime})=&g_{\mu\nu}(x)\frac{\partial x^{\mu}}{\partial x^{\prime \mu^{\prime}}}\frac{\partial x^{\nu}}{\partial x^{\prime \nu^{\prime}}}\cr
=&g_{\mu\nu}(x)[\delta^{\mu}_{\mu^{\prime}}-\frac{\partial \epsilon^{\mu}}{\partial x^{\prime \mu^{\prime}}}]
[\delta^{\nu}_{\nu^{\prime}}-\frac{\partial \epsilon^{\nu}}{\partial x^{\prime \nu^{\prime}}}]~~~.\cr
\end{align}
Treating $\epsilon$ as a first order perturbation, dropping second order terms, and using the fact that because of spatial homogeneity of the
RW metric the zeroth order metric has no dependence on the difference between $x^{\prime}$ and $x$, Eq. \eqref{gauge2} reduces to a gauge
transformation formula for the difference $\delta_gh_{\mu \nu}\equiv g^{\prime}_{\mu\nu}-g_{\mu\nu}$~~~,
\begin{align}\label{metricgauge}
\delta_g h_{ij}=&a^2(t) (\partial_j\epsilon^i+\partial_i\epsilon^j)~~~,\cr
\delta_g h_{i0}=&a^2(t)\partial_0\epsilon ^i~~~~,\cr
\delta_g h_{00}=&0~~~.\cr
\end{align}
Since the zeroth order particulate matter stress-energy tensor is also spatially homogeneous, the same
reasoning applies to calculating the gauge variation of the first order particulate matter stress-energy tensor, and we find using Eq. \eqref{zerostress} that (see \cite{wein} for further details)
\begin{equation}\label{stressengauge}
\delta_g T_{\mu\nu{\rm pm}}^{(1)}=-p\,\delta_g h_{\mu \nu}
\end{equation}
Comparing this with $\delta_g$ applied to  Eq. \eqref{stresspert}, we learn that the perturbed quantities $p^{(1)}$, $u^{(1)}$, $\rho^{(1)}$, $\pi^S$, $\pi_i^V$, $\pi_{ij}^T$, $u_i^V$ are all invariant under the gauge transformation of Eq. \eqref{metricgauge}.

Let us now calculate the variations of the first order Ricci tensor components given in Eq. \eqref{riccitensor} under the gauge transformation of Eq.
\eqref{metricgauge}.  After a lengthy calculation, in which many terms cancel,  we find for the ${\mu\nu}={ij},{i0},{00}$ cases
\begin{equation}\label{riccigauge}
\delta_g  R_{\mu\nu}=[\frac{\ddot{a}}{a}+2(\frac{\dot{a}}{a})^2]\delta_g h_{\mu\nu}~~~.
\end{equation}
Since $\delta_g t_{\mu\nu}=0$, the gauge variation of Eq. \eqref{ricciform} becomes, using Eq. \eqref{zerothordercosm}
\begin{equation}\label{ricciform1}
\delta_g[R^{(1)}_{\mu \nu}-\Lambda h_{\mu\nu}]=[\frac{\ddot{a}}{a}+2(\frac{\dot{a}}{a})^2-\Lambda]\delta_g h_{\mu\nu}
 =8\pi G\frac{1}{2}(\rho-p)\delta_g h_{\mu\nu}=-8 \pi G \delta_g[T_{\mu \nu{\rm pm}}-\frac{1}{2} g_{\mu\nu} T^\alpha_{\alpha {\rm pm}}]^{(1)}~~~.
\end{equation}
Thus gauge invariance of the first order perturbation equations requires
\begin{equation}\label{gaugeident}
\frac{1}{2}(p-\rho)\delta_g h_{\mu\nu}=\delta_g[T_{\mu \nu{\rm pm}}-\frac{1}{2} g_{\mu\nu} T^\alpha_{\alpha {\rm pm}}]^{(1)}~~~.
\end{equation}
In Appendix C we show that Eq. \eqref{gaugeident} is implied by Eq. \eqref{stressengauge}, and so the first order perturbation equations are
gauge invariant.

Having established gauge invariance of the perturbed equations, we are free to make a choice of gauge to simplify the subsequent calculations.
Taking
\begin{equation}\label{bfix}
\epsilon^i=\frac{1}{2} \partial_i B,
\end{equation}
we get $\delta_g h_{ij}= a^2 \partial_i \partial_j B$, which cancels the $B$ term in Eq.  \eqref{split}, giving what one might
term ``restricted Newtonian gauge''.  However, one cannot also gauge $F$ to zero as in full Newtonian gauge, since this requires use
of $\epsilon_0$.  We also cannot attain synchronous gauge, since this again requires  use of $\epsilon_0$.

\section{The modified scalar perturbation equations}

Combining Eqs. \eqref{zerothordercosm}, \eqref{firstorder1}, \eqref{split}, \eqref{riccitensor}, and \eqref{gaugeident1}, and choosing $B=0$ gauge, we get the following results
for the modified scalar perturbation equations.   They correspond to Eqs. (5.1.44)-(5.1.47) of Weinberg \cite{wein}, with the omission of the $B$ terms,
and the addition of the $\Lambda f t_{\mu\nu}$ terms arising from the frame dependent effective action.\footnote{Note that $\Lambda$ terms not multiplied by $f$ have cancelled when Eqs. \eqref{zerothordercosm} and \eqref{gaugeident1} were used. That is why our $f=0$ equations in $B=0$ gauge are identical to those
of \cite{wein}, which omits a cosmological constant.} The $ij$ part of the scalar perturbation can be written as
\begin{align}\label{ijpart}
0=&\delta_{ij}X+\partial_i\partial_jY~~~,\cr
X=&4\pi G a^2(\rho^{(1)}-p^{(1)}-\nabla^2 \pi^S)\cr
+&[a \ddot{a} + 2(\dot{a})^2]E+\frac{1}{2}a \dot{a}\dot{E}+\dot{a}\nabla^2F-3a\dot{a}\dot{A}-\frac{1}{2}a^2\ddot{A}+\frac{1}{2}\nabla^2 A\cr
+&\Lambda f a^2 (\frac{1}{2} t_{00}- E)~~~,\cr
Y=&8\pi G a^2 \pi^S\cr
+&2\dot{a}F+a\dot{F}+\frac{1}{2}(E+A)~~~.\cr
\end{align}
The $i0$ part of the scalar perturbation is
\begin{align}\label{i0part}
0=&-8 \pi G (p+\rho) \partial_iu^{(1)}\cr
- &\frac{\dot{a}}{a}\partial_iE +\partial_i\dot{A}\cr
+&\Lambda f t_{0i}~~~,\cr
\end{align}
and the $00$ part of the scalar perturbation is
\begin{align}\label{00part}
0=&4 \pi G(\rho^{(1)}+3p^{(1)}+\nabla^2\pi^S)\cr
-&3\frac{\ddot{a}}{a}E+3\frac{\dot{a}}{a}\dot{A}+\frac{3}{2}\ddot{A}-\frac{3}{2}\frac{\dot{a}}{a} \dot{E}-\frac{\dot{a}}{a^2}\nabla^2 F
-\frac{1}{a}  \nabla^2\dot{F}-\frac{1}{2a^2} \nabla^2 E\cr
+&\Lambda f (\frac{1}{2}t_{00}+3E)~~~.\cr
\end{align}
We have displayed these equations with separate lines giving the matter perturbation source terms, the metric terms, and the additional pieces proportional to
$\Lambda f t_{\mu \nu}$.  The latter are given, we recall, by Eq. \eqref{integrate1} of Sec. 2, with $h_{00}=E$.

Since the particulate matter stress-energy tensor $T_{\mu \nu{\rm pm}}$ and the added stress-energy tensor  $t_{\mu\nu}$ are separately covariantly conserved, $D^{\mu}T_{\mu \nu{\rm pm}}=0$, $D^{\mu}t_{\mu\nu}=0$, the momentum and energy conservation equations given in  Eqs. (5.1.48) and (5.1.49) of \cite{wein} are
unmodified.  Also, since $t_{\mu\nu}$ involves only the scalar $h_{00}$,  the vector and tensor perturbation equations are unmodified.

\section{Absence of propagating scalar gravitational waves}

We turn now to the question of whether the modified scalar perturbation equations of the preceding section admit propagating  wave solutions.
Thus, we investigate whether the homogeneous equations, obtained by dropping the matter source terms, have solutions.  Since these equations have
space-independent coefficients, we can Fourier analyze with respect to the coordinate $\vec x$, and it suffices to keep a generic mode
$e^{i\vec k \cdot \vec x}$.  Once this is done, the $X$ and $Y$ terms in Eq. \eqref{ijpart} decouple, since the tensors $\delta_{ij}$ and $k_ik_j$ are
linearly independent.  So we are left with {\it four} coupled equations to solve.  We can now anticipate the answer, before doing detailed
arithmetic:  Since gauge invariance, which was used to set $B=0$,  has reduced the number of unknowns  from four to the set of {\it three} comprising  $A$, $F$, and $E$, the homogeneous equations are overdetermined.  So unless there is a hidden linear dependence (which we shall see there is not) there are no solutions, other than the trivial solution $A=F=E=0$.  This
turns out to be the case independent of the coefficient $f$ of the added term $\Lambda f t_{\mu\nu}$.

To start our calculation, we rescale $F \to F/a$, after which all terms in the scalar perturbation equations with equal numbers of spatial
derivatives have the same power of $a$ as coefficient.
Since the gravitational waves of interest for binary pulsars or black hole mergers have much shorter periods than the Hubble time scale
on which $a(t)$ changes, we make the approximation of treating $a$, $\dot{a}$, and $\ddot{a}$ as constants.  The homogeneous scalar equations then
become differential equations for the evolution of $A$, $F$, and $E$ with constant coefficients, which we solve by making an Ansatz of $e^{-i\omega t}$
time dependence (where $\omega$ has a small negative imaginary part so the wave vanishes at $t=-\infty$) and $e^{i\vec k \cdot a\vec x}$ spatial dependence (with $a$ treated here as time-independent).\footnote{Regarding the $a$ in the spatial wave as $a(t_0)$ for some late time $t_0$, our approximation thus
consists of neglecting terms of order $H(t-t_0)$, $[H(t-t_0)]^2$, etc. in the Taylor expansion of the coefficient functions $a(t)$, $\dot{a}(t)$, etc.
at $t=t_0$.  When applied to the tensor perturbation equation of Eq. (5.1.53) of \cite{wein}, this procedure leads to the equation
$\nabla^2 D_{ij}-\ddot{D}_{ij}-3H \dot{D}_{ij}=0$, which describes a weakly damped propagating wave.}
We write  $\dot{a}\equiv H a$, $\ddot{a} \equiv H^2 Q a$, with constant $H$, $Q$, and $a$, replace $\partial_j$ by $iak_j$,  $\partial_t$ by $-i\omega$, and $\int dt$ by $(-i\omega)^{-1}$, and factor away
the uniform powers of $a$ in all terms of the equations (which is equivalent to setting $a=1$).
The time integrations of Eq. \eqref{integrate1} now give
\begin{align}\label{teqs}
t_{i 0}=& -2 \frac{k_i}{\omega}E~~~,\cr
t_{00}=&\left[2\frac{(\vec k)^2}{\omega^2}-6i\frac{H}{\omega}\right]E~~~\cr
\end{align}
Using these, we then  get the following
set of four coupled equations for the now constant unknowns $A$, $F$, and $E$:  From $Y$
in the $ij$ equation of Eq. \eqref{ijpart} we get
\begin{equation}\label{ypart}
0=HF-i\omega F+\frac{1}{2}(E+A)~~~;
\end{equation}
from the $i0$ equation of Eq. \eqref{i0part} we get (after
factoring out $k_i$),
\begin{equation}\label{i0part1}
0=-iHE+\omega A-\frac{2\Lambda f}{\omega}E~~~;
\end{equation}
from $X$ in the $ij$ equation of  Eq. \eqref{ijpart} we get
\begin{equation}\label{xpart}
0=(Q+2)H^2E-\frac{1}{2}i\omega H E-(\vec k)^2 H F+3i\omega H A+\frac{1}{2}\omega^2 A-\frac{1}{2} (\vec k)^2 A
+\Lambda f \left[\frac{(\vec k)^2}{\omega^2}-3i\frac{H}{\omega}-1\right]E~~~;
\end{equation}
and from the $00$ equation of Eq. \eqref{00part} we get
\begin{equation}\label{00part1}
0=-3QH^2 E-3i\omega H A-\frac{3}{2}\omega^2 A+\frac{3}{2}i\omega H E
-i\omega (\vec k)^2 F+\frac{1}{2}(\vec k)^2 E+\Lambda f \left[\frac{(\vec k)^2}{\omega^2}-3i\frac{H}{\omega}+3\right]E~~~.
\end{equation}

Eqs. \eqref{ypart} and \eqref{i0part1} can be solved for $F$ and $A$ in terms of $E$, giving
\begin{align}\label{fesoln}
F=&\frac{1}{2i\omega} \frac{1+\frac{iH}{\omega}+\frac{2\Lambda f}{\omega^2}}{1+\frac{iH}{\omega}} E~~~,\cr
A=&\left(\frac{iH}{\omega}+\frac{2\Lambda f}{\omega^2}\right) E~~~.\cr
\end{align}
Substituting these into Eqs. \eqref{xpart} and \eqref{00part1}, we can write the results respectively in the form
\begin{align}\label{results}
0=&E[(\vec k)^2 \alpha(\omega) + \beta(\omega)]~~~,\cr
0=&E[(\vec k)^2 \gamma(\omega) + \delta(\omega)]~~~,\cr
\end{align}
with
\begin{align}\label{alphetc}
\alpha(\omega)=&\gamma(\omega)= (\frac{i\Lambda f H}{\omega^3})/(1+\frac{iH}{\omega})~~~~,\cr
\beta(\omega)=&(Q-1)H^2+3\frac{i\Lambda fH}{\omega}~~~,\cr
\delta(\omega)=&-3\beta(\omega)~~~.\cr
\end{align}
Evidently the two equations in Eq. \eqref{results} are inconsistent, and so can only be solved by $E=0$.

One caveat to this analysis is that the algebra leading to Eqs. \eqref{alphetc} involves multiple cancellations
of leading terms, and so there could be significant corrections when the time dependence of $a$, $\dot{a}$, $\ddot{a}$ is
taken into account.  But there is no reason for these corrections to conspire to make the two equations of Eq. \eqref{results}
consistent.

\section{Conclusion}

We have shown that  cosmological action of Eq. \eqref{ansatz} has potentially testable consequences for the equations governing scalar
perturbations around the RW metric.  This may make it possible to distinguish between the standard and the frame dependent actions, both of which
give rise to an effective cosmological constant.

\section{Acknowledgements}
I wish to thank a number of people for raising the question of whether the frame-dependent effective action affects gravitational wave propagation:
Clifford Burgess and Andrei Barvinsky (at the Galiano Island workshop), Edward Witten (in conversations) and Jeremy Bernstein (in emails).  I also wish
to thank Matias Zaldarriaga for email correspondence calling my attention to the effective field theory framework of \cite{gub}, and to thank
Paul Steinhardt for a helpful conversation.
\appendix
\section{Notational conventions}

Since many different notational conventions are in use for gravitation and cosmology, we summarize here
the notational conventions used in this paper and in \cite{adler1}-\cite{adler3}.

(1)~~ The Lagrangian in flat spacetime is $L=T-V$, with $T$ the kinetic energy and $V$ the potential energy, and
the flat spacetime Hamiltonian is $H=T+V$.

(2)~~ We use a $(1,-1,-1,-1)$ metric convention, so that in flat spacetime, where the metric is denoted by $\eta_{\mu \nu}$, the various 00 components of
the stress energy tensor $T_{\mu \nu}$ are equal, $T_{00}=T_{0}^{0}=T^{00}$.

(3)~~The affine connection, curvature tensor, contracted curvatures, and the Einstein tensor, are
given by
\begin{align}
\Gamma^{\lambda}_{\mu \nu} =&\frac{1}{2}g^{\lambda \sigma}(g_{\sigma \nu,\, \mu}+g_{\sigma \mu,\, \nu}- g_{\mu \nu,\, \sigma})~~~,\cr
R^{\lambda}_{~\mu \tau \nu}=& \Gamma^{\lambda}_{~\mu \tau,\, \nu}-\Gamma^{\lambda}_{~\mu \nu,\, \tau} +{\rm quadratic ~terms ~in~} \Gamma  ~~~,\cr
R_{\mu \nu}=&R^{\lambda}_{~\mu \lambda \nu}=\Gamma^{\lambda}_{\mu \lambda, \, \nu}-\Gamma^{\lambda}_{\mu \nu, \, \lambda}+{\rm quadratic~terms~in~}\Gamma~~~,\cr
R=&g^{\mu \nu} R_{\mu \nu}~~~,\cr
G_{\mu \nu}=&R_{\mu \nu}-\frac{1}{2} g_{\mu \nu} R ~~~.\cr
\end{align}

(4)~~ The Einstein-Hilbert gravitational action and its variation with respect to the metric $g_{\mu \nu}$ are
\begin{align}\label{eq:lambda0}
S_{\rm EH} = & \frac{1}{16\pi G} \int d^4x (^{(4)}g)^{1/2} R~~~,\cr
\delta S_{\rm EH}=& -\frac{1 }{16 \pi G} \int d^4x (^{(4)}g)^{1/2} G^{\mu \nu}\delta g_{\mu \nu}~~~.\cr
\end{align}

(5)~~The particulate matter action and its variation with respect to the metric $g_{\mu \nu}$ are
\begin{align}\label{eq:matteraction}
S_{\rm pm}=& \int dt  L = \int d^4 x (^{(4)}g)^{1/2}  {\cal L}(x)~~~,\cr
\delta S_{\rm pm}=&-\frac{1}{2} \int d^4 x (^{(4)}g)^{1/2}  T^{\mu \nu}_{\rm pm}  \delta g_{\mu \nu}~~~.\cr
\end{align}

(6)~~The Einstein equations with cosmological constant $\Lambda$ are
\begin{equation}\label{ricciform0}
G^{\mu \nu}+ \Lambda g^{\mu \nu} +8 \pi G T^{\mu \nu}_{\rm pm}+\Lambda  f t^{\mu \nu}=0~~~,
\end{equation}
with the final term the additional term arising from the frame dependent effective action,
which we have split off by writing  $ T_\Lambda^{\mu\nu}=\frac{\Lambda}{8\pi G}(g^{\mu\nu}+ f t^{\mu \nu})$.
Eq. \eqref{ricciform0} can equivalently be written as
\begin{equation}\label{ricciform}
R_{\mu \nu}-\Lambda g_{\mu\nu}=-8 \pi G [T_{\mu \nu{\rm pm}}-\frac{1}{2} g_{\mu\nu} T^\alpha_{\alpha {\rm pm}}]
- \Lambda f[ t_{\mu \nu}-\frac{1}{2} g_{\mu\nu}^{(0)}  t^\alpha_{\alpha }]~~~,
\end{equation}
with $g_{\mu\nu}^{(0)}$ the  unperturbed RW metric.

(7)  To compare with Weinberg \cite{wein}, our metric $g_{\mu\nu}$ and metric perturbation $h_{\mu\nu}$ are opposite in
sign to his, while our relations between  the Ricci tensor $R_{\mu \nu}$, the affine connection $\Gamma^\lambda_{\mu \nu}$, and the
metric are the same.   Since the affine connection is an even function of the metric, this means that the zeroth order parts of the
affine connection and the Ricci tensor are the same as in \cite{wein}, while the first order metric perturbations in these quantities
are opposite in sign.  In defining the scalar, vector, and tensor parts of the metric perturbations, we introduce an extra minus sign
relative to \cite{wein},
\begin{align}\label{split}
h_{00}=&E~~~,\cr
h_{i0}=&-a(\partial_iF+G_i)~~~,\cr
h_{ij}=&-a^2(A\delta_{ij}+\partial_i\partial_jB+\partial_jC_i+\partial_iC_j+D_{ij})~~~,
\end{align}
with $\partial_iC_i=\partial_iG_i=\partial_iD_{ij}=D_{ii}=0$.  The quantities $E$, $F$, $G_i$, $A$, $B$, $C_i$, $D_{ij}$ then are
the same as in \cite{wein}.  Our energy momentum tensor sign is the same as in Weinberg,  but reflecting our opposite sign of $h_{\mu\nu}$
we define the tensorial decomposition of the perturbed stress-energy tensor   $T^{(1)}_{\mu\nu {\rm pm}}$ by
\begin{align}\label{stresspert}
T^{(1)}_{ij {\rm pm}}=&-p\,h_{ij}+a^2[\delta_{ij}p^{(1)}+\partial_i\partial_j\pi^S+\partial_i\pi_j^V+\partial_j\pi_i^V+\pi_{ij}^T]~~~,\cr
T^{(1)}_{i0 {\rm pm}}=&-p\,h_{i0}-(p+\rho)(\partial_iu^{(1)}+u_i^V)~~~,\cr
T^{(1)}_{00 {\rm pm}}=&\rho\, h_{00} +\rho^{(1)}~~~,\cr
\end{align}
with $\partial_i\pi_i^V=\partial_iu_i^V=\partial_i\pi_{ij}^T=\pi_{ii}^T=0$.  The quantities $p^{(1)}$, $u ^{(1)}$,  $\rho^{(1)}$, and $u_i^V$  are then
the same as Weinberg's $\delta p$, $\delta u$, $\delta \rho$, and $\delta u_i^V$, and the quantities $\pi^S$, $\pi_i^V$, and $\pi_{ij}^T$ are the
same as his similarly labeled quantities.

\section{Inverse metric, affine connections, and Ricci curvature tensor for the perturbed RW metric}

The inverse perturbed RW metric corresponding to Eq. \eqref{totalmetric} is given, through first order terms,
by
\begin{align}\label{totalinversemetric}
g^{00}=&1-h_{00}~~~,\cr
g^{i0}=&g^{0i}=h_{i0}/a^2(t)~~~,\cr
g^{ij}=&-\delta_{ij}/a^2(t)-h_{ij}/a^4(t)~~~.\cr
\end{align}
The perturbed RW affine connection, through first order terms, is
\begin{align}\label{affineconn}
\Gamma^0_{00}=&\frac{1}{2}\partial_0 h_{00}~~~,\cr
\Gamma^0_{0i}=&\frac{1}{2}\partial_i h_{00}-\frac{\dot{a}}{a}h_{0i}~~~,\cr
\Gamma^0_{ij}=&a\dot{a}(1-h_{00})\delta_{ij}+\frac{1}{2}(\partial_jh_{0i}+\partial_ih_{0j}-\partial_0h_{ij})~~~,\cr
\Gamma^{\ell}_{00}=&-\frac{1}{a^2}(\partial_0h_{\ell 0}-\frac{1}{2}\partial_{\ell}h_{00})~~~,\cr
\Gamma^{\ell}_{0i}=&\frac{\dot{a}}{a}\delta_{i\ell}-\frac{1}{2a^2}(\partial_0 h_{\ell i}+\partial_i h_{\ell 0}-\partial_{\ell}h_{0i})
+\frac{\dot{a}}{a^3}h_{\ell i}~~~,\cr
\Gamma^{\ell}_{ij}=&\frac{\dot{a}}{a}h_{\ell 0} \delta_{ij}-\frac{1}{2a^2}(\partial_jh_{\ell i}+\partial_ih_{\ell j}-\partial_{\ell}h_{ij})~~~.\cr
\end{align}
The corresponding Ricci tensor components are
\begin{align}\label{riccitensor}
R_{00}=&3\frac{\ddot{a}}{a}+\frac{\dot{a}}{a^3}\partial_0h_{mm}+\frac{1}{a^2}[\frac{\ddot{a}}{a}-(\frac{\dot{a}}{a})^2]h_{mm}
-\frac{3}{2}\frac{\dot{a}}{a}\partial_0h_{00}+\frac{1}{a^2}\partial_0\partial_mh_{m0}-\frac{1}{2a^2}(\partial_0^2h_{mm}+\nabla^2h_{00})~~~,\cr
R_{0i}=&[\frac{\ddot{a}}{a}+2(\frac{\dot{a}}{a})^2]h_{i0}-\frac{\dot{a}}{a}\partial_ih_{00}+\frac{\dot{a}}{a^3}
[\partial_ih_{mm}-\partial_mh_{mi}]+\frac{1}{2a^2}(\partial_m\partial_0h_{mi}+\partial_i\partial_mh_{m0}-\nabla^2h_{0i}-\partial_0\partial_ih_{mm})~~~,\cr
R_{ij}=&-\delta_{ij}[a\ddot{a}+2(\dot{a})^2](1-h_{00})+\delta_{ij}[\frac{1}{2}a\dot{a}\partial_0h_{00}-\frac{\dot{a}}{a}\partial_mh_{m0}
+\frac{1}{2}\frac{\dot{a}}{a}\partial_0h_{mm}-(\frac{\dot{a}}{a})^2h_{mm}]\cr
-&\frac{1}{2}(\partial_0\partial_jh_{0i}+\partial_0\partial_ih_{0j}
-\partial_0^2h_{ij}-\partial_i\partial_jh_{00})+\frac{1}{2a^2}(\partial_m\partial_jh_{mi}+\partial_m\partial_ih_{mj}
-\nabla^2h_{ij}-\partial_i\partial_jh_{mm})\cr-&\frac{\dot{a}}{2a}(\partial_0h_{ij}+\partial_jh_{0i}+\partial_ih_{0j})
+2(\frac{\dot{a}}{a})^2h_{ij}~~~.\cr
\end{align}

\section{Equivalence of two forms of the gauge variation of the first order particulate matter stress-energy
tensor}

We proceed to show that Eq. \eqref{gaugeident} is implied by Eq. \eqref{stressengauge}.  Expanding
\begin{align}\label{expand}
[T_{\mu \nu{\rm pm}}-\frac{1}{2} g_{\mu\nu} T^\alpha_{\alpha {\rm pm}}]^{(1)}
=&T_{\mu \nu{\rm pm}}^{(1)}-\frac{1}{2} h_{\mu\nu}T^{\alpha(0)}_{\alpha {\rm pm}}-
\frac{1}{2} g_{\mu\nu}^{(0)} T^{\alpha(1)}_{\alpha {\rm pm}}~~~,\cr
 T^{\alpha(1)}_{\alpha {\rm pm}}=&g^{\mu\nu(0)} T^{(1)}_{\mu\nu {\rm pm}}+g^{\mu\nu(1)} T^{(0)}_{\mu\nu {\rm pm}}~~~,\cr
\end{align}
and substituting the decompositions of Eq. \eqref{stresspert}, we get
\begin{align}\label{gaugeident1}
 T^{\alpha(1)}_{\alpha {\rm pm}}=&\rho^{(1)}-3p^{(1)}-\nabla^2 \pi^S~~~,\cr
[T_{00{\rm pm}}-\frac{1}{2} g_{00} T^\alpha_{\alpha {\rm pm}}]^{(1)}=&
\frac{1}{2}(3p+\rho)h_{00}+\frac{1}{2}(\rho^{(1)}+3p^{(1)}+\nabla^2 \pi^S)~~~,\cr
[T_{0i{\rm pm}}-\frac{1}{2} g_{0i} T^\alpha_{\alpha {\rm pm}}]^{(1)}=&
\frac{1}{2}(p-\rho)h_{0i}-(p+\rho)(\partial_iu^{(1)}+u_i^V)~~~,\cr
[T_{ij{\rm pm}}-\frac{1}{2} g_{ij} T^\alpha_{\alpha {\rm pm}}]^{(1)}=&
\frac{1}{2}(p-\rho)h_{ij}+a^2\delta_{ij}\frac{1}{2}(\rho^{(1)}-p^{(1)}-\nabla^2 \pi^S)+a^2(\partial_i\partial_j \pi^S+\partial_i\pi_j^V+\partial_j\pi_i^V+\pi_{ij}^T)~~~.\cr
\end{align}
Applying $\delta_g$ to these equations, and using $0=\delta_g h_{00}=\delta_g\rho^{(1)}=\delta_gp^{(1)}=\delta_g u^{(1)}=\delta_g\pi^S=
\delta_g \pi_i^V=\delta_g\pi_{ij}^T=\delta_gu_i^V$, we obtain
Eq. \eqref{gaugeident}.    The detailed decompositions of Eq. \eqref{gaugeident1} are used in writing down the detailed form of the scalar
perturbation equations in Sec. 4.

\section{Relation between our dark energy model and the effective field theory of dark energy \cite{gub}}

We show here that although our model can be mapped to the framework of \cite{gub}, the
underlying physics is different.  In particular, their method cannot be used to test the frame-dependent  model because their
procedure leads to a non-covariantly conserved dark energy stress-energy tensor, as we shall now
show in detail.

When the action of Eq. \eqref{effact3} is expanded to first order in perturbations by writing $g_{00}=1+h_{00}$, it takes the form
\begin{align} \label{ansatz1}
S_{\rm eff}
=& -\frac{\Lambda}{8\pi G} \int d^4x (^{(4)}g)^{1/2} [1-2h_{00}]\cr
=&-\frac{\Lambda}{8\pi G} \int d^4x (^{(4)}g)^{1/2} [-1+2g^{00}]\cr
\end{align}
where on the second line we have substituted (to first order accuracy) $g^{00}=1-h_{00}$.
Taking $8\pi G=1$, and comparing with Eq. (1) of \cite{gub}, noting that Eq. \eqref{ansatz1}  is written in the
Einstein frame, and that the sign convention for the metric of \cite{gub} is opposite to ours, we get the following identifications
of their coefficient functions,
\begin{align}\label{gubparam}
f(t)=&1~~{\rm (Einstein~ frame)}~~~,\cr
\Lambda(t)= &-\Lambda~~~\cr
c(t)=&-2\Lambda~~~,\cr
\end{align}
that is, their three coefficient functions are all {\it constants}, independent of time in our model.

Up to this point our model has fit into the framework of \cite{gub} ; the difference shows up at their Eq. (14), where they vary
with respect to the full $g^{\mu\nu}$  to get the Einstein equations.  Since we have $f=1$, their
Eq. (14) becomes
\begin{equation}\label{eineq}
G_{\mu\nu} M_{\star}^2+(c\delta g^{00}+\Lambda-c)g_{\mu\nu} -2c\delta^{0}_{\mu}\delta^{0}_{\nu}=T_{\mu\nu}~~~.
\end{equation}
With constant $c$ and $\Lambda$, this equation is inconsistent at zeroth order in perturbations, since $g_{\mu\nu}$, $G_{\mu\nu}$ and $T_{\mu\nu}$ are covariantly
conserved on the RW background, while the term $c\delta^{0}_{\mu}\delta^{0}_{\mu}$ is not covariantly conserved.
To see this explicitly, switching to the upper index version of this equation we have
\begin{align}
D_{\nu}c \delta^{\mu}_0\delta^{\nu}_0=&\partial_{\nu}c \delta^{\mu}_0\delta^{\nu}_0
+\Gamma^{\mu}_{\nu\alpha}c \delta^{\alpha}_0\delta^{\nu}_0+ \Gamma^{\nu}_{\nu\alpha} c \delta^{\mu}_0\delta^{\alpha}_0\cr
=&[\dot{c}+ 3 \frac{\dot{a}}{a} c]\delta^{\mu}_0~~~,\cr
\end{align}
where in the second line we have substituted the RW affine connection $\Gamma^{\ell}_{0i}=\frac{\dot{a}}{a}\delta_{i\ell}$.
We see that when $c$ is a constant, so $\dot{c}=0$, covariant conservation fails.  This problem is a direct reflection
of the fact that our model is not four-space diffeomorphism invariant. This problem does not arise in the paper \cite{gub}, because their
$c(t)$, $\Lambda(t)$ etc. are obtained by transforming an underlying model that {\it is} diffeomorphism invariant  to unitary
gauge, leading to non-constant $c(t)$, $\Lambda(t)$ etc. which take values that guarantee covariant conservation of their dark
energy stress-energy tensor.

For our frame-dependent model, the correct way to get a covariantly conserved dark energy stress-energy tensor is given in Sec. 2  above.   First one varies
with resect to the spatial components $g_{ij}$ to generate the spatial components $T^{ij}$ of the stress-energy tensor; then one integrates the
covariant conservation equations to get the remaining components $T^{i0}$ and $T^{00}$. This procedure works both because the action of Eq. \eqref{ansatz1} is three-space general coordinate invariant, and because in the ADM formulation \cite{adm} of general relativity, the $g_{ij}$ are the fundamental degrees of freedom of the gravitational
field.

\section{Off-diagonal metric terms in the effective action do not change the first order perturbation equations}
When the metric has nonzero off-diagonal terms, as is the case for RW perturbations, the effective action of Eq. \eqref{effact2} generalizes
\cite{adler1} to
\begin{equation}\label{geneffact}
S_{\rm eff}= \int d^4x (^{(4)}g)^{1/2} (g_{00})^{-2}A(h_{0i}h_{0j}g^{ij}/g_{00}, D^ig_{ij}D^j/g_{00},h_{0i}D^i/g_{00})~~~,
\end{equation}
with $D^i$ defined through the co-factor expansion of $^{(4)}g$ by $^{(4)}g/^{(3)}g=g_{00}+h_{0i}D^i$,  with $A(x,y,z)$ a general function of
its arguments, and where we have used $g_{0i}=h_{0i}$.   Evaluating $D^i$ from its definition, we find
\begin{align}\label{evalD}
D^1=&-\frac{h_{10}}{g_{11}}+O(h^2)~~~,\cr
D^2=&-\frac{h_{20}}{g_{22}}+O(h^2)~~~,\cr
D^3=&-\frac{h_{30}}{g_{33}}+O(h^2)~~~.\cr
\end{align}
From this we see that $D^i$ and $\delta D^i/\delta g_{kl}$ are both $O(h)$, as a consequence of which
\begin{equation}
\delta A(h_{0i}h_{0j}g^{ij}/g_{00}, D^ig_{ij}D^j/g_{00},h_{0i}D^i/g_{00})/\delta g_{kl}=O(h^2)~~~.
\end{equation}
Thus when we vary $S_{\rm eff}$ with respect to the spatial metric components $g_{kl}$, the additions to $T_{\Lambda}^{kl}$ coming from
the nonzero arguments of the function $A$ are second order corrections to the first order result of Eq. \eqref{vary2}.  Hence in a first order
calculation, it suffices to take $A$ as $A_0=A(0,0,0)$, as was done in the sections above.

\end{document}